# Are polar liquids less simple?

D. Fragiadakis and C.M. Roland
Naval Research Laboratory, Chemistry Division, Code 6120, Washington DC 20375-5342

*- Aug 28, 2012 -*

**ABSTRACT** Strong correlation between equilibrium fluctuations of the potential energy, *U*, and the virial, *W*, is a characteristic of a liquid that implies the presence of certain dynamic properties, such as density scaling of the relaxation times and isochronal superpositioning of the relaxation function. In this work we employ molecular dynamics simulations (mds) on methanol and two variations, lacking hydrogen bonds and a dipole moment, to assess the connection between the correlation of *U* and *W* and these dynamic properties. We show, in accord with prior results of others [T. S. Ingebrigtsen, T.B. Schrøder, J.C. Dyre, Phys. Rev. X **2**, 011011 (2012).], that simple van der Waals liquids exhibit both strong correlations and the expected dynamic behavior. However, for polar liquids this correspondence breaks down – weaker correlation between *U* and *W* is not associated with worse conformance to density scaling or isochronal superpositioning. The reason for this is that strong correlation between *U* and *W* only requires their proportionality, whereas the expected dynamic behavior depends primarily on constancy of the proportionality constant for all state points. For hydrogen-bonded liquids, neither strong correlation nor adherence to the dynamic properties is observed; however, this nonconformance is not directly related to the concentration of hydrogen bonds, but rather to the greater deviation of the intermolecular potential from an inverse power law (IPL). Only (hypothetical) liquids having interactions governed strictly by an IPL are perfectly correlating and exhibit the consequent dynamic properties over all thermodynamic conditions.

**INTRODUCTION**

Classic studies of the liquid state address “simple liquids”, commonly defined as molecules interacting via spherically-symmetric pair potentials having additive attractive and repulsive energies [1]. Such terms are loosely defined, and simplicity can refer to an absence of quantum effects, while “complex liquids” may include large, bulky molecules or substances having anisotropic interactions or coexisting phases. Supercooled liquids are associated with many-bodied interactions, and thus can be viewed as simple liquids that have complex dynamics. Recently, Dyre and coworkers [2]developed the idea of “strongly correlating liquids” as a prototype for simple liquids, especially relevant to the viscous, dense state. A strongly correlating liquid is defined by strong correlations between equilibrium fluctuations in the virial, *W*, and potential, *U*, energies; such correlations are perfect (i.e., proportionality of *W* and *U* with an invariant proportionality constant) for particles interacting according to an inverse power law potential [3]. An important feature of this approach is that when liquids are shown to be strongly correlating, they are expected to exhibit other characteristic properties [2]. One such property is density scaling, whereby dynamic quantities are a function of the ratio of temperature to density with the latter raised to a material constant $\gamma$

$$X = f\left(T / \rho^{\gamma}\right) \tag{1}$$

In eq. (1) *f* is a function and $X$ represents relaxation time $\tau$, viscosity $\eta$, or diffusion constant *D,* expressed in reduced units. Another property expected for correlating liquids is isochronal superpositioning of their relaxation function, meaning the distribution of relaxation times is constant for any fixed value of $\tau$. Simulations have found that correlating liquids have intermolecular potentials dominated by van der Waals interactions, while hydrogen-bonded materials show deviations in the correlation of *W* and *U* [4]. The properties of density scaling and isochronal superpositioning have been found to follow accordingly [5,6].

It is expected that molecular associations engendered by hydrogen bonding markedly affect the dynamics of liquids [6,7,8,9,10]. Research has focused mainly on water, which is the most ubiquitous and important H-bonded material. The viscosity of water is larger than that of other liquids of similar molecular weight, due to the presence of H-bonds, and reorientation of water molecules requires coordinated switching of a hydrogen atom between different H-bond partners [11,12]. However, H-bonding in water is atypical, being associated with an extended network structure that imposes specific distances and molecular orientations, due to the ability of one water molecule to form multiple H-bonds. These factors cause the concentration of H-bonds in water to *decrease* with pressure as the network is disturbed [13,14,15]. This reduction in H-bonding enhances molecular motions, countervailing the direct effect of densification on the dynamics. Whether such effects are unique to water is an open question. Some experiments [16,17,18,19,20] and mds [21,22] have shown that pressure *increases* H-bonding in non-aqueous liquids, while other studies have found that in alcohols, pressure either weakens hydrogen bonds [7,23] or has no effect [24].

H-bonded liquids share some dynamical properties with water, such as isochoric-isobaric activation enthalpy ratios for the viscosity that approach unity [25,26] and a breakdown of density scaling [9]; nevertheless, the unique structure of water makes it problematic as a model to interpret the dynamics of hydrogen bonded liquids. In the present work we carry out molecular dynamics simulations on a simple compound able to form hydrogen bonds. Our molecules consist of a methyl group with either a hydroxyl or oxygen moiety (respectively, methanol or a non-H-bonded version having the same dipole moment), or the latter sans any dipole moment. This variation of the chemical structure systematic alters the capacity for intermolecular associations. We study the effect of thermodynamic variables on these associations and on the dynamics, and evaluate the applicability of the correlating liquids approach to materials having hydrogen bonds or polar interactions.

## EXPERIMENTAL

Three structures were studied:

(i) **MeOH**, a rigid, united-atom model of methanol taken from the GROMOS force field [27]. This is the same model previously used in developing the pressure-energy correlation hypothesis [28,29]. The potential energy combines Lennard-Jones (6-12) and Coulomb interactions

$$U(r) = 4\varepsilon_{ij}\left[\left(\frac{\sigma_{ij}}{r}\right)^{12} - \left(\frac{\sigma_{ij}}{r}\right)^{6}\right] + \frac{q_i q_j}{4\pi\varepsilon_0 r} \qquad (2)$$

with three sites representing methyl (M), oxygen (O), and hydroxyl hydrogen (H), having respective masses *m* = 15.035, 15.999, and 1.008 Da. Lennard-Jones parameters are $\varepsilon_{MM}$=0.9444, $\varepsilon_{OO}$=0.8496, and $\varepsilon_{MO}$ =0.9770 kJ/mol, and $\sigma_{MM}$=0.3646, $\sigma_{OO}$=0.2955, and $\sigma_{MO}$=0.3235 nm. Charges are 0.176 *e*, -0.574 *e*, 0.398 *e* for the M, O, and H sites, respectively. *M*–O and O–H distances are fixed respectively at 0.136 and 0.1 nm, while the *M*–O–H bond angle is fixed at 108.53°.

(ii) **MeO** is a two-site molecule with the same dipole moment as MeOH, but with no H atom, thus precluding H-bonding [30,31]. Molecular weight is equal to that of MeOH, with *m* = 15.035 and 17.007 Da for the two sites. The Lennard-Jones interactions are also set equal to those of MeOH. Charges are 0.290 *e* and -0.290 *e*, with the M-O distance fixed at 0.136 nm.

(iii) ***n*MeO** is a two-site molecule identical to MeO but without charges and thus having zero Coulomb interactions.

Simulations were carried out using GROMACS [32,33,34]. Bond lengths and angles were maintained constant using the LINCS algorithm [35]. Lennard-Jones interactions are smoothly truncated between 0.9 and 1.2 nm. Electrostatic interactions are calculated using the particle mesh Ewald method [36]. Temperature and pressure are controlled using the Nosé-Hoover thermostat [37] and isotropic Parinello-Rahman barostat [38,39], respectively. For each system, *N* = 1728 molecules were simulated, with the simulations carried out at pressures up to 20 GPa, at several temperatures for each pressure. At each state point an *NPT* run was performed for equilibration, signified by no significant drift in volume or aging in the translational and rotational correlation functions, followed by an *NVT* run during which data were collected.

In mds different criteria have been proposed to identify H-bonds, falling generally into two categories: energetic and geometric. We use the geometric criterion introduced by Haughney et al. [40]: distances $d_{OO}$≤ 0.35nm and $d_{OH}$≤ 0.26nm, and angle $\vartheta_{OH\cdots O}$ ≤ 30°; all three must be satisfied. In the lower alcohols the manner of defining an H-bond has been shown to have little effect on the computed results [40,41].

**RESULTS**

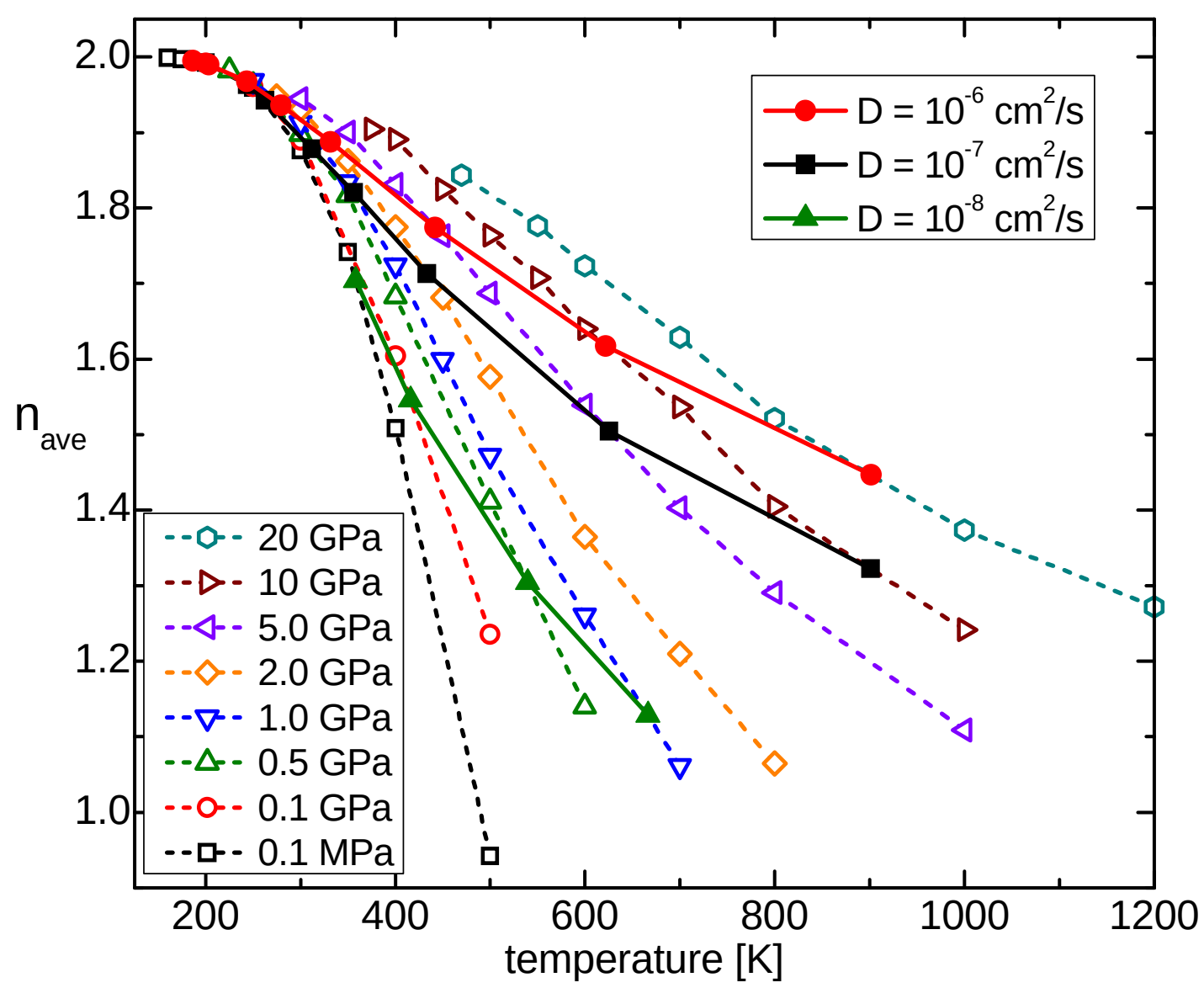

**Figure 1.** Average number of H-bonds per MeOH molecule as a function of temperature along the indicated isobars (open symbols). Filled symbols indicate curves of constant self-diffusion coefficient.

In agreement with previous results [16,19], the fraction of MeOH molecules having hydrogen bonds decreases with temperature and increases with pressure. The majority engage in two H-bonds, but this shifts to more singly-bonded species (terminal H-bonds) at higher *T*, while higher pressure causes an increase of doubly and triply bonded species. **Figure 1** shows the average number of H-bonds per molecule ($n_{ave}$) as a function of *T* and *P*. The effect of pressure is amplified at higher temperatures, while at low *T* (200-300K), $n_{ave} \sim 2$ independent of *P*. It is interesting to examine hydrogen bonding properties at constant dynamics, wherein the effects of *T* and *P* mutually compensate. The self-diffusion coefficient *D* was calculated from the slope of the mean-square displacement using

$$D = \lim_{r\to\infty} \frac{\langle r^2(t) \rangle}{6t} \tag{3}$$

Included in Fig. 1 is $n_{ave}$ at constant value of *D*. The behavior along these curves is dominated by temperature, so that there is substantially less H-bonding at higher (*T,P*) for constant *D*. Results (not shown) were equivalent for constant relaxation time. This demonstrates that the degree of H-bonding is not the dominant control variable.

The presence of H-bonds and dipoles does influence the dynamics, as seen in **Figure 2** comparing *D* for the three species (H-bonded MeOH, polar MeO and neutral *n*MeO). The densities for the three systems cover a similar range, for each system varying by a factor of ~2 (much more than in a typical experiment). At higher pressures the mobilities are essentially the same, even though for MeOH $n_{ave}$ is close to 2. Evidently at very high pressures jamming and molecular packing effects, which are similar for the three species, overwhelm the contribution of the coulombic interactions (see discussion below).

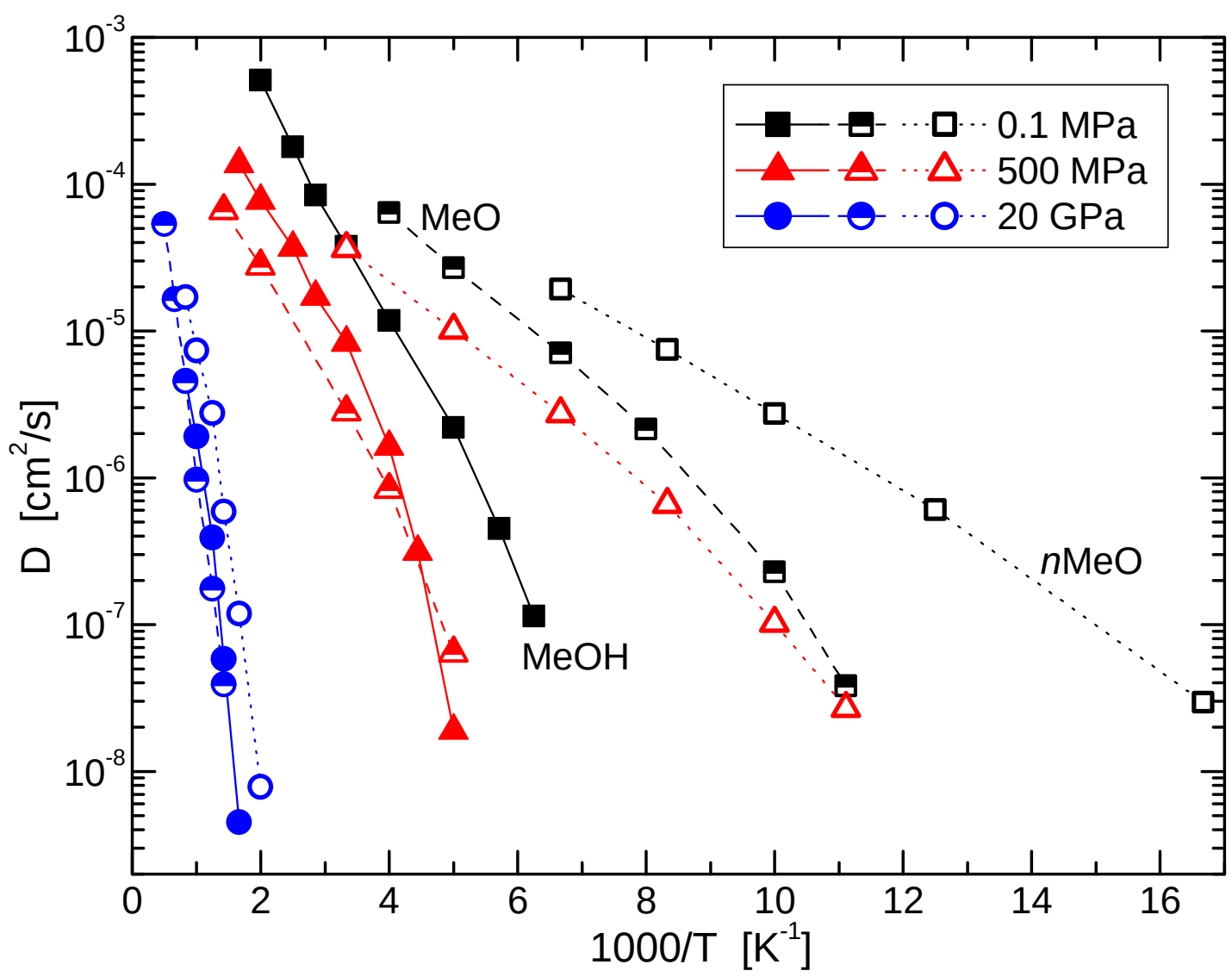

**Figure 2**. Arrhenius plot of self-diffusion coefficient at various pressures for MeOH (filled symbols), MeO (half-filled symbols), and *n*MeO (open symbols).

**Pressure-Energy Correlations**

Dyre and coworkers [2,5] proposed that strong correlation (linear correlation coefficient $R \geq 0.9$) between equilibrium fluctuations of the virial and potential energy defines a simple liquid. Using this definition they found that associated molecular liquids are not strongly correlating and thus not simple, except at very high densities [2,28]. As shown in Table 1, listing the *R* values for each system herein averaged over all state points, the correlation between *W* and *U* degrades going from *n*MeO > MeO > MeOH, consistent with these previous results. For MeOH *R* has a wide range of values for different conditions of *T* and *P*, with the correlation generally improving as the system is compressed. However, the generally poor correlation for MeOH cannot be ascribed directly to hydrogen-bonding, as evident from the data in **Figure 3**. Moreover, the *W-U* correlation is also poor for MeO (which lacks H-bonds) except at the highest densities. H-bonding *per se* does not underlie departures from *W-U* correlation.

Plotted in **Figure 4** is *R* vs the ratio of the magnitudes of the Coulomb and L-J forces, averaged over all molecules. MeOH becomes a correlating liquid at higher densities for which the van der Waals forces become predominant over the coulombic interactions from both H-bonded and polar interactions. Interestingly, the data in Fig. 4 fall on a single curve (except for deviations at low P where the attractive part of the Lennard-Jones potential becomes significant). These results underscore that it is not H-bonding, but the presence of substantial coulombic forces that degrades the connection between *W* and *U*. This is consistent with the fact that the property of correlation is strictly adhered to only for IPL potentials [3]; the Coulomb term in eq. (2) makes this a poorer approximation.

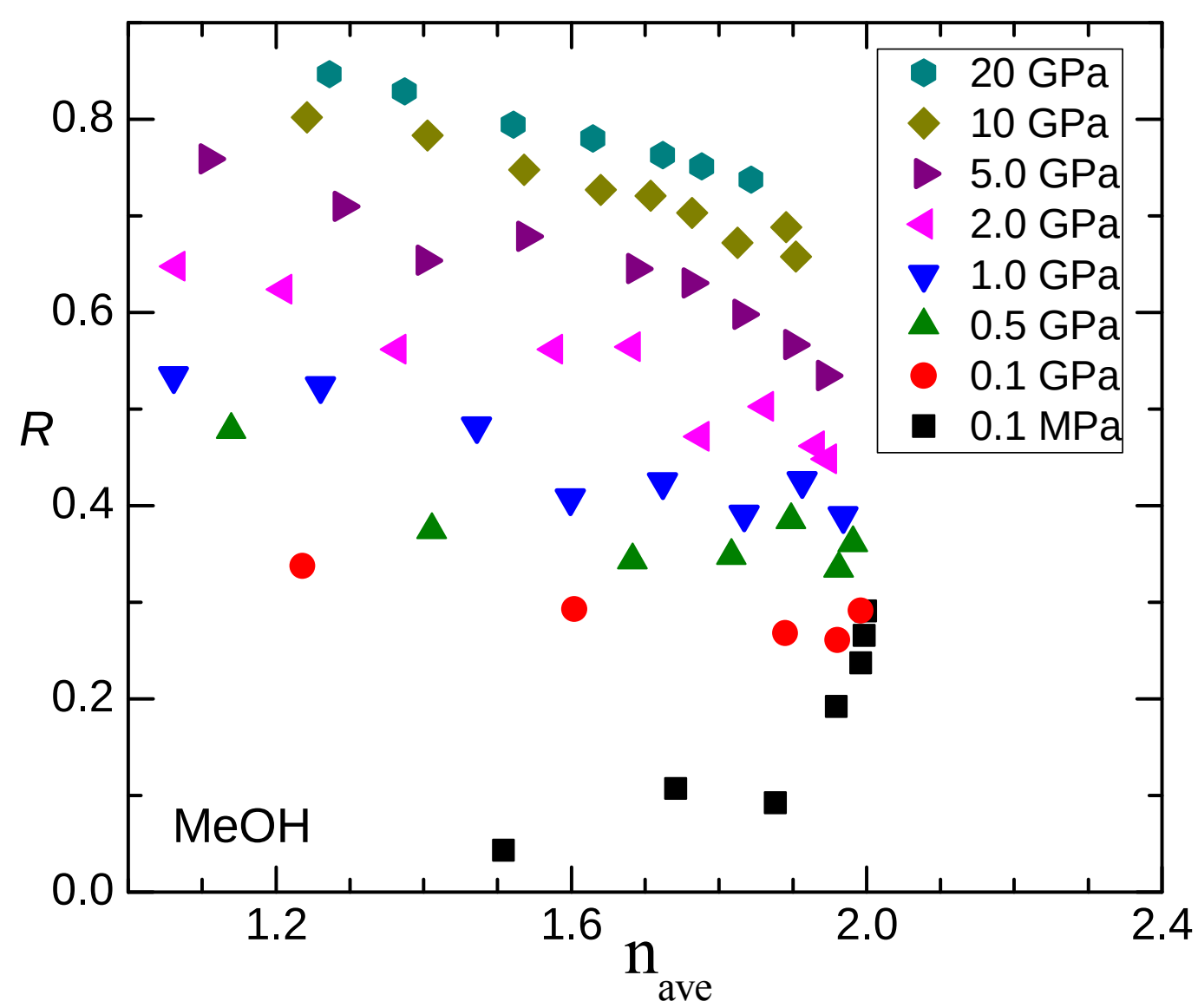

**Figure 3.** Correlation coefficient for virial – potential energy correlations vs. average number of H-bonds per methanol molecule. Each data point corresponds to a different state point (P, T).

Table 1. Correlation coefficient and proportionality constant for W-U fluctuations

| | $R_{avr}$ | *dW/dU* | $\gamma$ *(slope)* | $\gamma$ *(scaling)* |
|---|---|---|---|---|
| *n*MeO | 0.97±0.01 | 5 – 6.4 | 5.5 | 5.7 |
| MeO | 0.87±0.07 | ~4.7 | 4.6 | 4.55 |
| MeOH | 0.52±0.19 | 0.3 – 4.0 | 2.3 | 2.9 |

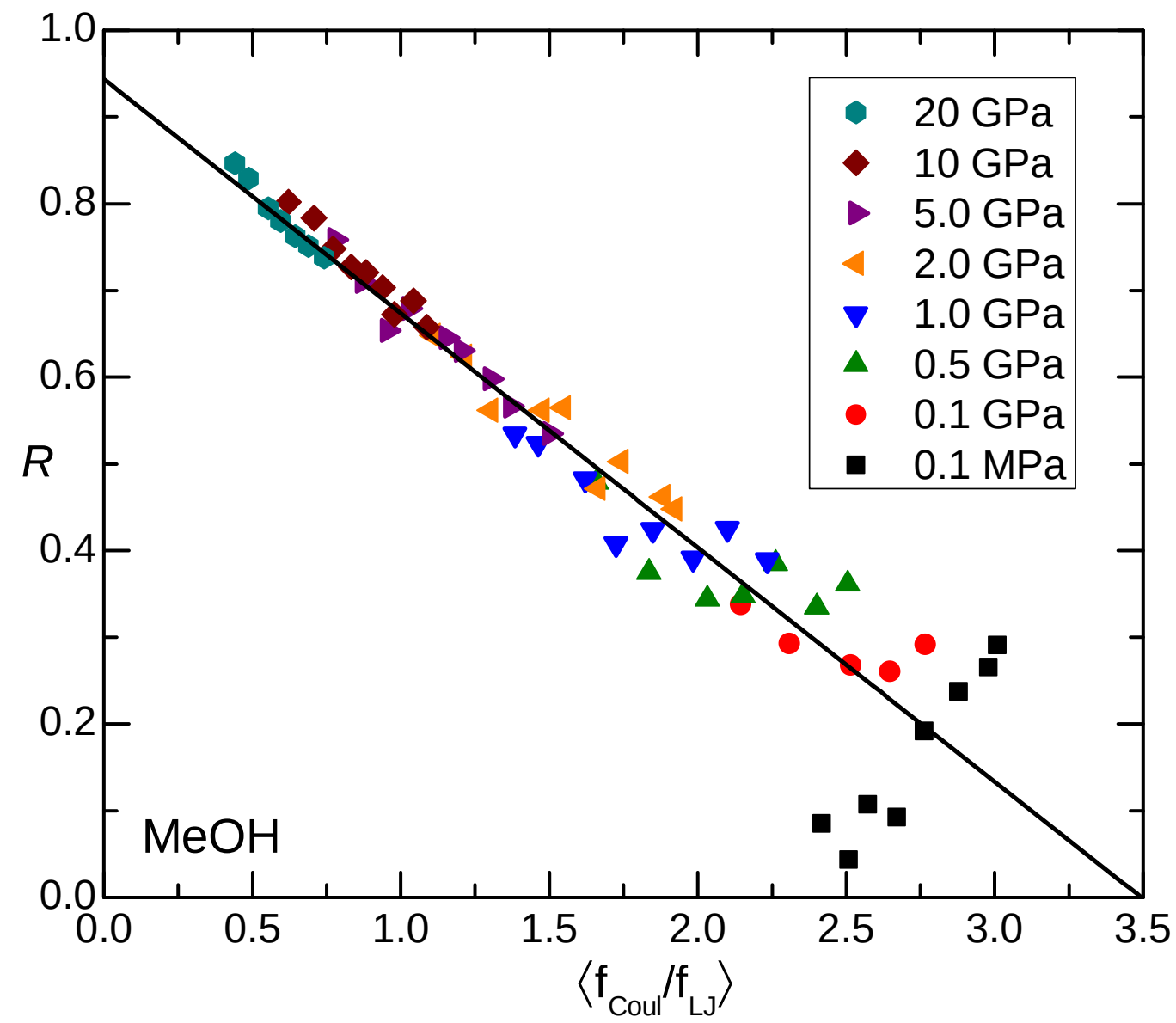

**Figure 4.** Pressure-energy correlation coefficient for methanol vs. the ratio of the magnitudes of Coulomb and Lennard-Jones forces on a molecule averaged over all molecules, for each measured state point.

Fig. 4 raises the question – why do van der Waals forces dominate at higher pressures? To examine this we plot in **Figure 5** the center-of-mass radial distribution function (rdf) for three state points at which the hydrogen bonding of MeOH is approximately constant ($n_{ave}$ = 1.60 ± 0.02). At low pressures there is a sharp peak due to H-bonded clusters at distances smaller than the average intermolecular separation; similar results have been reported previously [30]. Beyond the second peak the rdf of the three liquids is nearly the same. However, with increasing pressure the peak due to the clusters merges with the first near-neighbor peak, as the structure becomes governed primarily by packing considerations; effects due to H-bonds are minified, as reflected in the decreasing value of $\left\langle f_{coul}/f_{LJ}\right\rangle$. Without a change in the *number* of H-bonds, the repulsive Lennard-Jones term becomes dominant as the intermolecular peak in the rdf supersedes the peak due to H-bonded associations. With diminution of the effect of the H-bonding on the intermolecular interactions, the fluctuations in *W* and *U* become more correlated.

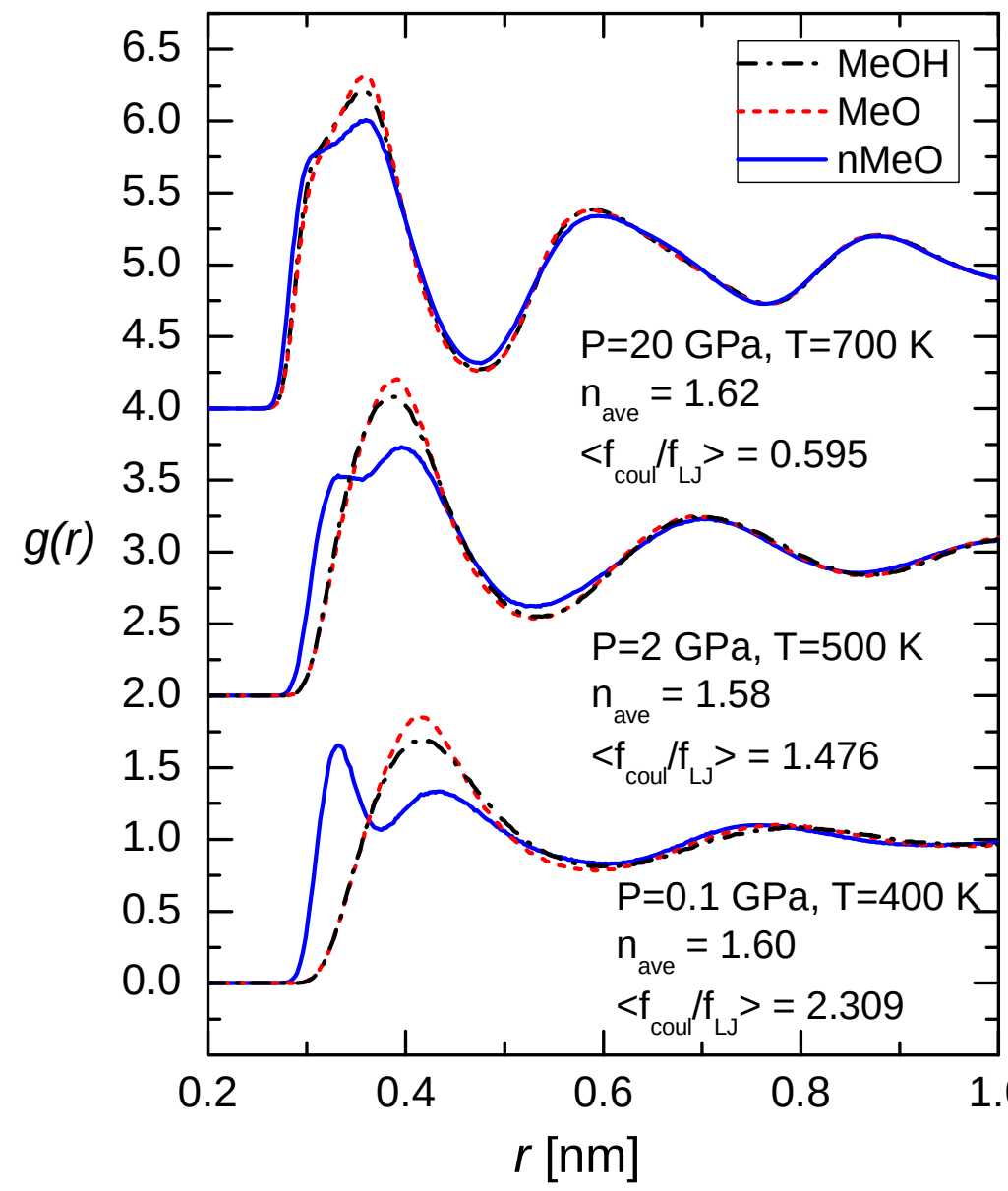

**Figure 5.** Center-of-mass radial distribution function at three state points for which $n_{ave}$ = 1.60 ± 0.02 for MeOH. On densification, the effect of the H-bonding is minimized and the Lennard-Jones interactions become dominant. (The curves for 2 and 20 GPa have been vertically shifted by 2 and 4, respectively).

**Density scaling**

A limitation of using the correlation of *W* and *U* as a means to classify liquids is the difficulty in experimentally assessing this correlation (although recently it was shown that the proportionality constant between fluctuations in *W* and *U* can be calculated from linear thermoviscoelastic data for a single state point [42]). However, the expectation that certain properties are integral to correlating liquids implies that the latter can be identified by adherence to these properties. Thus, MeOH and MeO would conform poorly to density scaling, reflecting their departures from strongly correlating behavior (Table 1). However, while H-bonded liquids exhibit poor density scaling [6], it is well established that dynamic data for polar liquids density scale quite accurately; indeed, many examples of density scaling are drawn from dielectric relaxation experiments that rely on molecular dipole moments for the measurements [25]. Even for ionic liquids, characterized by very substantial coulombic forces, viscosities have been shown to density scale [9,43].

A useful means to assess density scaling is from double logarithmic plots of temperature vs. density for constant values of, for example, the relaxation timescale. Thus, from eq. (1) a power law form is predicted

$$T\rho^{-\gamma}\Big|_{\tau,\eta,D} = \text{constant} \tag{4}$$

with a constant slope equal to the scaling exponent $\gamma$. **Figure 6** shows for a constant value of the diffusion constant (using reduced values, $D^* = \rho^{1/3}\left(kT/m\right)^{-1/2} D$ [44]), *T* vs. $\rho$ for our three liquids. As expected the data for MeOH deviate from the linearity expected for scaling. The plot for *n*MeO is closer to linear, although interestingly the scaling is best for MeO. This result, inconsistent with the relative degrees of *W*-*U* correlation exhibited by the two liquids lacking H-bonds, has its explanation in the fact that only an IPL potential yields behavior complying exactly with eq.(4). An L-J potential is well approximated by an IPL only in the region near the minimum in *U(r)*, and for this reason the scaling behavior deviates significantly at lower densities. For polar liquids, the Coulomb term in eq. (2) increases the steepness of the potential near the minimum, counteracting the effect of the attractive interactions. The result is that although there is poorer correlation between *W* and *U*, the proportionality constant between these quantities remains nearly constant (equal to the slope of the repulsive part of the potential near the minimum [45]). Thus, even though *n*MeO is more strongly correlating ($R \geq 0.94$ for every state point), MeO conforms better to density scaling because there is a less variation of $\gamma$ as thermodynamic conditions are changed. The strongly correlating property of *n*MeO does not yield especially good scaling because $\gamma$ differs significantly for different state points.

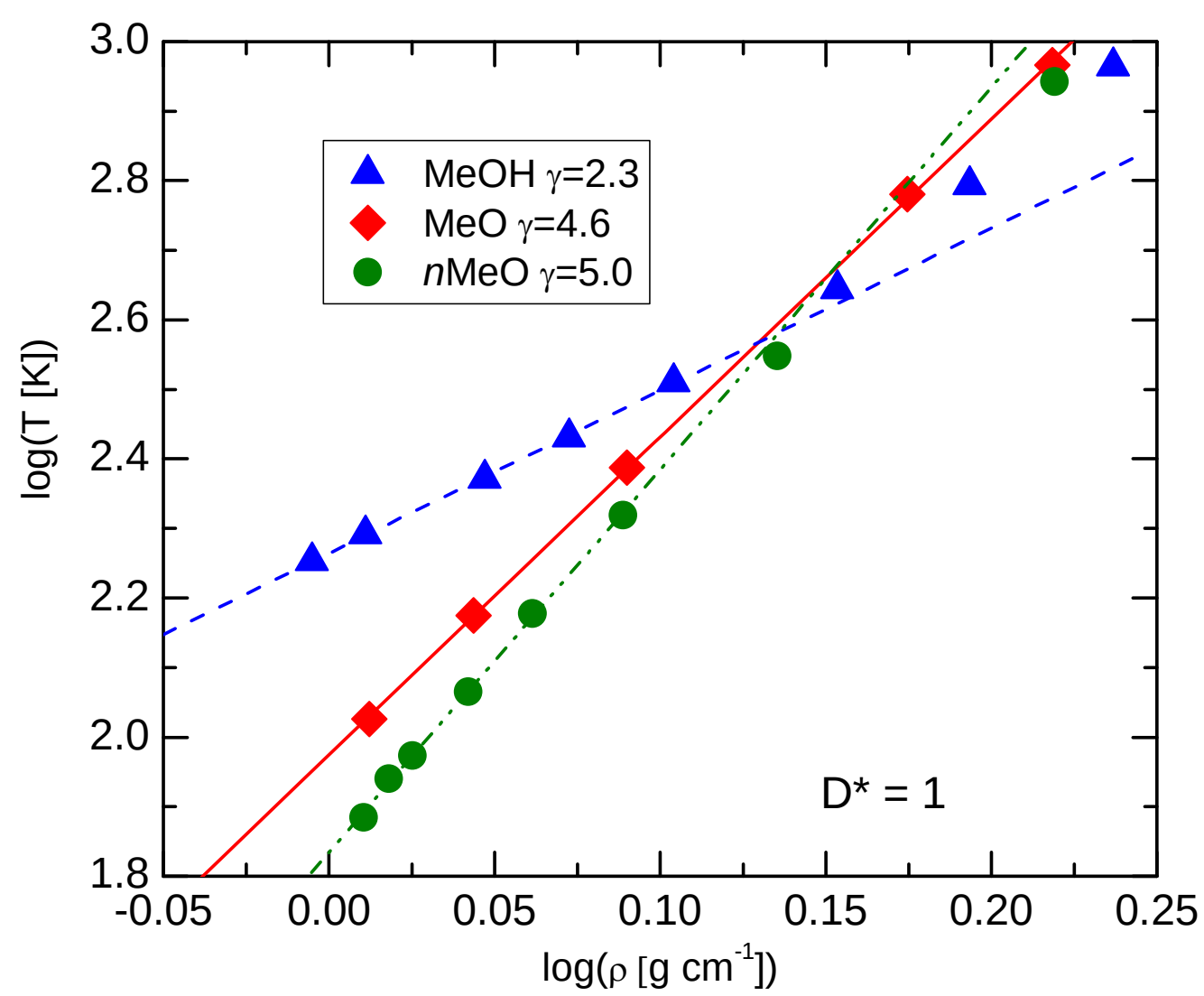

**Figure 6.** Temperature vs. density at the indicated constant value of the reduced diffusion constant. The lines represent power law fits (eq.(4)) for the lower densities; only for MeO does the fit describe the data for $\rho > 0.1$.

To compare the scaling behavior from the mds to experimental results, the range of densities covered should be similar. For MeO in Fig. 6, $\rho$ varies by a factor of two. In dielectric spectroscopy experiments, density changes exceeding 20% are rare, although for some liquids experimental viscosities and diffusion coefficients are available over a much wider volume range and have been found to conform to density scaling [46]. Limiting the range of densities for the mds data to no more than a 40% higher variation in $\rho$, we obtain the density scaling plots of *D** shown in **Figure 7**. The data for both *n*MeO and MeO superpose well, in accord with the behavior of real molecular liquids. These results demonstrate that the correlating liquid concept can provide a useful interpretation of the behavior of real materials, provided the mds probe realistic thermodynamic conditions. That is, while density scaling breaks down for a sufficient change of densities, eq. (1) is valid under the ranges of *T* and *P* usually studied experimentally. For this reason the strength of the correlation of *W* and *U* is not reflected in the accuracy of density scaling for real materials (leaving aside the problem of measuring this correlation experimentally).

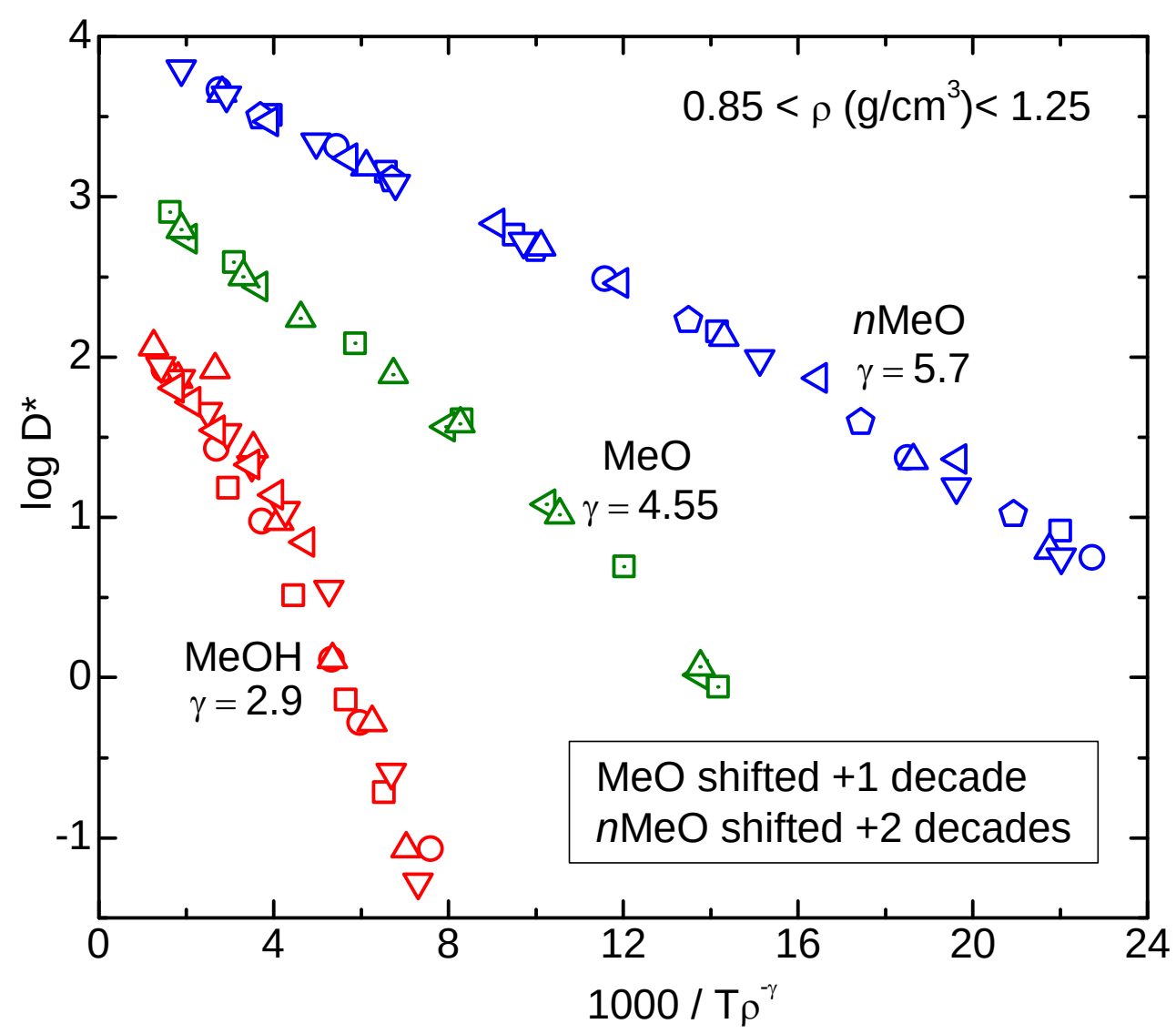

**Figure 7.** Reduced diffusion coefficient as a function of $1000/T\rho^{\gamma}$, where $\gamma$ is the empirical value that best collapses the data for each system. Each data point corresponds to a unique state point, with data for MeO and *n*MeO shifted vertically for clarity.

**Isochronal superposition**

Another property observed experimentally [47,48] that is expected for strongly correlating liquids is constancy of the relaxation function at fixed $\tau$. To assess this we fit a stretched exponential equation

$$\Phi(t) = \Phi_0 \exp\left(-\left(t/\tau\right)^{\beta_K}\right) \tag{5}$$

to the α relaxation (for MeOH we use the dipole autocorrelation function, while for *n*MeO we use the the 1st order rotational correlation function; for MeO the two functions are same). In **Figure 8** the obtained $\beta_K$ are plotted vs. temperature for different fixed pressures. Fojr MeOH at low *P* (<1 GPa), $\beta_K$ is large (~0.8) and relatively independent of temperature and pressure, so that isochronal superpositioning is trivial. For higher *P* the relaxation broadens, increasingly so as $\tau$ increases; the relaxation function is not fixed at constant $\tau$. For both liquids without H-bonding, isochronal superpositioning holds over the entire range of conditions. The relaxation is slightly more stretched in *n*MeO, and becomes broader with increasing $\tau$ for both systems. For MeO the poorer correlation of *W* and *U* is not reflected in inferior isochronal superpositioning.

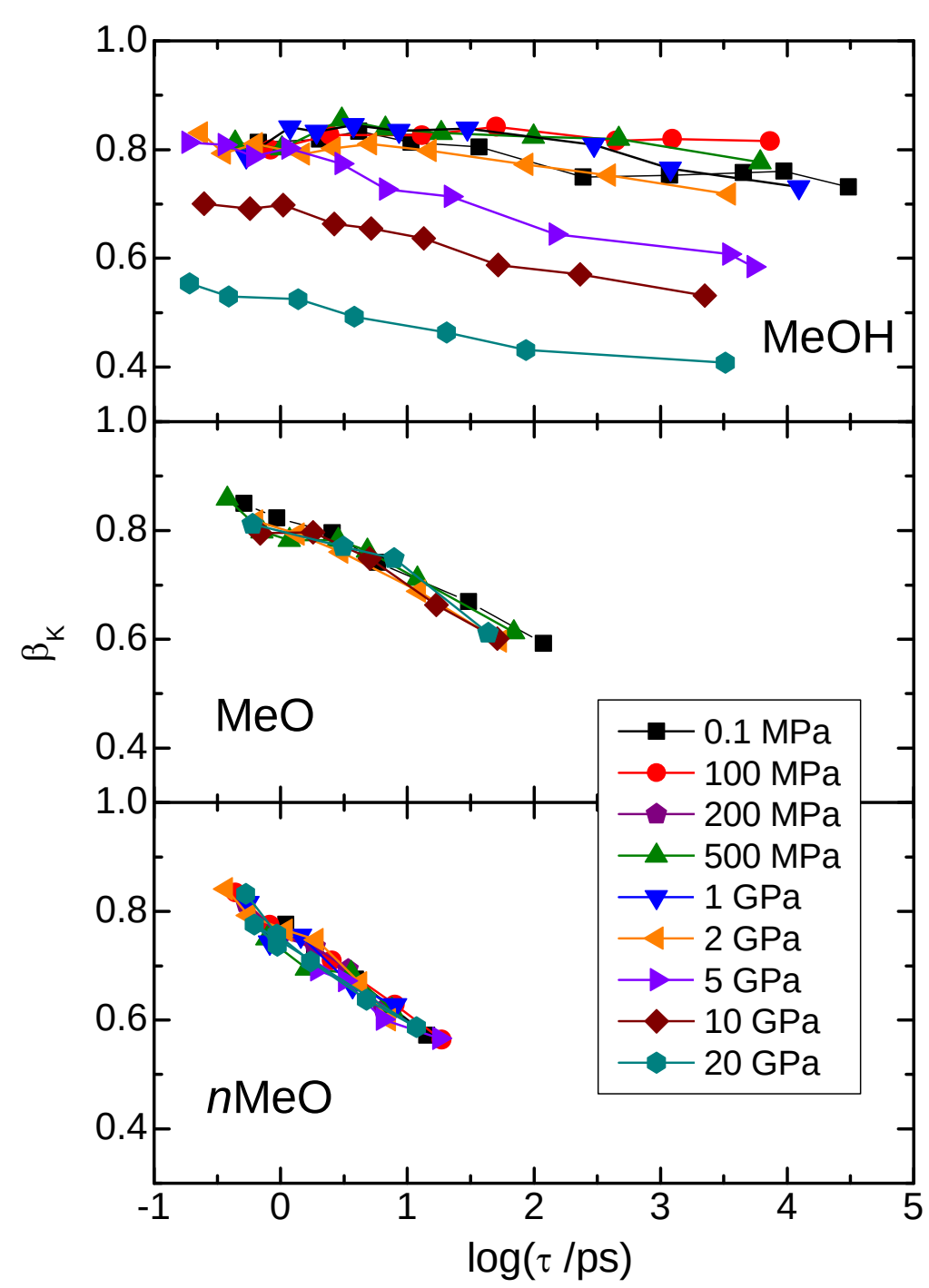

**Figure 8**. Stretch exponent from eq. (5) of the the 1st order rotational correlation function for *n*MeO and the dipole autocorrelation function for MeOH and MeO.

**SUMMARY**

Our main findings herein are as follows:

(i) At fixed temperature, pressure increases the extent of hydrogen-bonding in methanol, with the effect apparent *inter alia* in the magnitude of the self-diffusion coefficient.

(ii) The hydrogen-bonded liquid and the liquid having dipolar interactions both show poorer correlation of *W* and *U* equilibrium fluctuations. The degradation of this correlation is greater for MeOH; however, the effect cannot be ascribed to any direct effect of hydrogen bonding.

(iii) The correlation of *W* and *U* requires that van der Waals forces dominate the rdf, which for strongly associated liquids transpires only at higher densities.

(iv) The properties of density scaling and isochronal superpositioning are exhibited by the two liquids that lack H-bonds. However, the quality of the scaling and superpositioning is not related to the degree of correlation of *W* and *U*, because strong correlation only requires proportionality between *W* and *U*, whereas scaling and superpositioning depend primarily on invariance of the value of the proportionality constant over all state points (even if there is only moderate correlation).

**Acknowledgement**

This work was supported by the Office of Naval Research.